# Science Yield of an Improved Wide Field Infrared Survey Telescope (WFIRST)

M. E. Levi[1], A. G. Kim[1], M. L. Lampton[2], and M. J. Sholl[2]

[1] Physics Division, Lawrence Berkeley National Laboratory, Berkeley CA 94720
[2] Space Sciences Laboratory, University of California, Berkeley CA 94720

## Abstract

The Astronomy and Astrophysics Decadal Survey's highest recommended space mission was a Wide-Field Infrared Survey Telescope (WFIRST) to efficiently conduct three kinds of studies: dark energy surveys, exoplanet surveys, and guest surveys. In this paper we illustrate four potential WFIRST payloads that accomplish these objectives and that fully utilize optical and technical advances made since the community input to the Decadal Survey. These improvements, developed by our group, are: unobscured 1.3 or 1.5 m apertures; simultaneous dual focal lengths delivering pixel scales of 0.18" for imaging and 0.38" or 0.45" for slitless spectroscopy; the use of a prism in converging light for slitless spectroscopy; and payload features that allow up to 270 days/year observing the Galactic Bulge. These factors combine to allow WFIRST payloads that provide improved survey rates compared to previous mission concepts. In this report we perform direct comparisons of survey speeds for constant survey depth using our optical and exposure-time tools previously developed for JDEM. We further compare these four alternative WFIRST configurations to JDEM-$\Omega$ and to the European Space Agency's Euclid mission, and to an alternative Euclid configuration making use of the lessons learned here that delivers performance approaching that of WFIRST. We find that the unobstructed pupil is a major benefit to weak lensing owing to its tighter point spread function, improved signal to noise, and higher resolved galaxy count. Using two simultaneous plate scales in a fully focal system is practical and simplifies the optical train, and the use of a prism in converging light offers improved throughput compared to a grism. We find that a 45° outer baffle cutoff angle, combined with fully articulated solar panels and K-band antenna, substantially increase the exoplanet yield. These findings were presented at the 217th AAS conference (M. Levi, et al., Jan 2011).

Subject heading: Astronomical Instrumentation

## 1. Introduction

The recently published Decadal Survey "New Worlds, New Horizons in Astronomy and Astrophysics" (NRC, 2010) prioritized ground and space based astronomy for the coming decade. Their highest recommendation for space was a near-infrared survey mission they named Wide Field Infrared Survey Telescope (WFIRST); it was based largely on the NASA DOE Joint Dark Energy Mission (JDEM) family, in particular JDEM-$\Omega$ (Gehrels 2010). WFIRST will operate for five to ten years and will combine the three leading dark energy investigations (weak lensing, baryon acoustic oscillations, and supernovae) with a microlensing exoplanet survey, and allow additional studies proposed by guest investigators.

In the following sections we briefly review the WFIRST requirements, and then contrast four new alternative payload concepts with the original JDEM-Ω design on which the Decadal recommendations were based, and to the Euclid mission and a variant of Euclid. In section 4 we detail some added features of these WFIRST alternative payloads. We believe that significant payload and mission improvements will result by adopting one of these alternative configurations, with little or no impact on cost, schedule, or risk.

## 2. WFIRST Requirements

The dark energy goals are to measure the expansion history of the universe and the growth of large-scale structure, to provide tight constraints on the equation of state of dark energy, and test the validity of general relativity. The exoplanet research would reveal a population of planets made visible by microlensing against the dense stellar background of the Galactic Bulge. Guest investigator time would also be made available for astronomers needing wide or deep exposures in the near infrared. In Table 1, we list the requirements posed by the Decadal Survey and its prioritization panel as they bear on these three broad objectives.

TABLE 1
Functional requirements from the Decadal Survey

| Objective | Requirement | Page Reference* |
|---|---|---|
| Weak Lensing | Image $2 \times 10^9$ galaxies | DS 7-17 |
| | Shapes and photo-z's for $\sim 10^9$ galaxies | SF 1-18 |
| | Photometry for photo-z's | SF 1-15 |
| | Spectroscopy of calibration sample | SF 1-15 |
| Baryon Acoustic | Spectra of $2 \times 10^8$ galaxies | DS 7-17 |
| Oscillations | Spectroscopic redshifts of $\sim 10^8$ galaxies | SF 1-18 |
| | Redshifts for 1<z<2 | SF 1-15 |
| | Restframe IR photometry | SF 1-18 |
| Supernovae | Measure & classify $\sim 2000$ SNe | DS 7-17 |
| | Near-IR photometry for z<1 | SF 1-15 |
| Exoplanet Microlensing | Galactic Bulge microlensing survey | DS 34, SF 4-32, PP 6-8 |
| Guest Observers | Broad range of objectives | PP 6-2, 6-8 |
| Payload Optics | 1.5m telescope | DS 7-17 |

Note—*DS=Decadal Survey;  SF=Science Frontiers;  PP=Program Prioritization EOS Panel.

## 3. Four WFIRST Payloads

The starting point for this development effort was the collection of JDEM payloads that were developed jointly by NASA GSFC and DOE LBNL (Sholl et al 2008; 2009), culminating in the JDEM-Ω concept (Gehrels 2010) that used HgCdTe (MCT) sensors throughout. Since then we have explored the benefits of adopting three innovations: an unobscured telescope pupil (Lampton et al. 2010 and references therein), a simultaneous dual focal length (Sholl et al. 2010, Content et al. 2010; Grange et al 2010; Sholl et al. 2011), and a slitless spectrometer prism in converging light (Sholl et al 2010). For WFIRST, we considered two alternative ways to perform imaging and wide field spectroscopy simultaneously. One way is to use a common sky field but divide the light with a dichroic beamsplitter, with shorter wavelengths sent to the imager and longer wavelengths passing through a prism and focal reducer. This path leads to WFIRST-A (1.5m aperture, 8x4 MCT imager, 4x2 MCT spectrometer) and C (1.5m aperture, 5x5 MCT imager, 2x2 MCT spectrometer), detailed below. The other way is to divide the surveys spatially, with adjoining fields on the sky, each spanning the full wavelength range: WFIRST-B has an 8x4 imager plus two 2x2 spectrometers, and D has a 5x5 imager plus a 2x2 spectrometer. All imagers have a 0.18" pixel scale. In B and D the bulkier Cassegrain optics leave less room within existing 4m launcher fairings, so we had to reduce the primary mirror size to 1.3m for options B and D. An on-axis pupil routinely leads to a payload whose rear-end optics (spectrometers, cameras etc) are located behind the primary mirror, yielding a payload that is long and slender, while an off-axis configuration leads to a more boxlike payload whose secondary/tertiary optics and focal planes stand alongside the telescope, not behind it. A related requirement is to provide a several-square-meter passive thermal radiator closely coupled to the focal planes. Our four payloads satisfy both requirements because the entire shaded side of each payload is available for a passive radiator, and the package dimensions can be accommodated by existing 4m diameter EELV fairings (Heetderks, 2010).

TABLE 2
Alternative Dark Energy Mission Payload Configurations

| Payload | WFIRST A | WFIRST B | WFIRST C | WFIRST D | JDEM Ω | Euclid VIS | Euclid NIR | Euclid-Plus VIS | Euclid-Plus NIR |
|---|---|---|---|---|---|---|---|---|---|
| **TELESCOPE** | | | | | | | | | |
| Aperture, m | 1.5 | 1.3 | 1.5 | 1.3 | 1.5 | 1.2 | | 1.2 | |
| obstructed? | No | No | No | No | Yes | Yes | | No | |
| Dichroic? | Yes | No | Yes | No | No | Yes | | Yes | |
| **IMAGER** | | | | | | | | | |
| FoV†, sqdeg | 0.336 | 0.336 | 0.262 | 0.262 | 0.252 | 0.466 | 0.466 | 0.466 | 0.466 |
| λ range, μm | 0.5-1.5 | 0.5-2.0 | 0.5-1.5 | 0.5-2.0 | 0.4-2.0 | 0.5-0.9 | 0.9-2.0 | 0.5-0.9 | 0.9-2.0 |
| Pixel, arcsec | 0.18 | 0.18 | 0.18 | 0.18 | 0.18 | 0.10 | 0.3 | 0.10 | 0.3 |
| WFE*, nm RMS | 52 | 40 | 22 | 46 | 45 | 64 | 118 | 64 | 118 |
| Sensor chips | 32MCT | 32MCT | 25MCT | 25MCT | 24MCT | 36CCD | 16MCT | 36CCD | 16MCT |
| imager pixels | 134M | 134M | 105M | 105M | 101M | 604M | 67M | 604M | 67M |
| **BAO SPECTR** | | | | | | | | | |
| FoV†, sqdeg | 0.336 | 0.336 | 0.282 | 0.282 | 0.532 | n/a | 0.466 | n/a | 0.466 |
| Pixel, arcsec | 0.36 | 0.36 | 0.45 | 0.45 | 0.37 | | 0.3 | | 0.3 |
| WFE*, nm RMS | 60 | 103 | 73 | 72 | 118 | | 118 | | 118 |
| Sensor chips | 8 MCT | 8 MCT | 4 MCT | 4 MCT | 12 MCT | | 16MCT | | 16MCT |
| Spectro pixels | 34M | 34M | 17M | 17M | 50M | | 67M | | 67M |
| Disperser | 1 prism | 2 prisms | 1 prism | 1 prism | 2 prisms | | 4 grisms | | 1 prism |
| λ range, μm | 1.5-2.0 | 1.1-2.0 | 1.5-2.0 | 1.1-2.0 | 1.1-2.0 | | 1.0-2.0 | | 1.0-2.0 |
| z range at Hα | 1.3-2.0 | 0.7-2.0 | 1.3-2.0 | 0.7-2.0 | 0.7-2.0 | | 0.7-2.0 | | 0.7-2.0 |
| **SN SPECTRO** | | | | | | | | | |
| FoV†, sqarcsec | 100 | 100 | 100 | 100 | slitless | n/a | | n/a | |
| λ range, μm | 0.4-2.0 | 0.4-2.0 | 0.4-2.0 | 0.4-2.0 | | | | | |

Notes— BAO SPECTR is a wide field slitless BAO spectrometer; SN SPECTRO is an IFU coupled to a low dispersion prism spectrometer. *WFE = system wavefront error. †FoV = active field of view.

In Table 2, we list the characteristics of the four WFIRST candidates developed here and compare them with three other dark energy mission concepts, JDEM-Ω (Gehrels 2010), the ESA Euclid mission (Duvet 2010; Wallner et al 2010), and "Euclid Plus" which is Euclid with an unobscured pupil and its four grisms replaced by one prism. To estimate the system wavefront errors we have doubled the theoretically optimized WFEs to account for manufacturing and alignment errors. For Euclid we doubled the theoretical WFE reported by Sanz (2010).

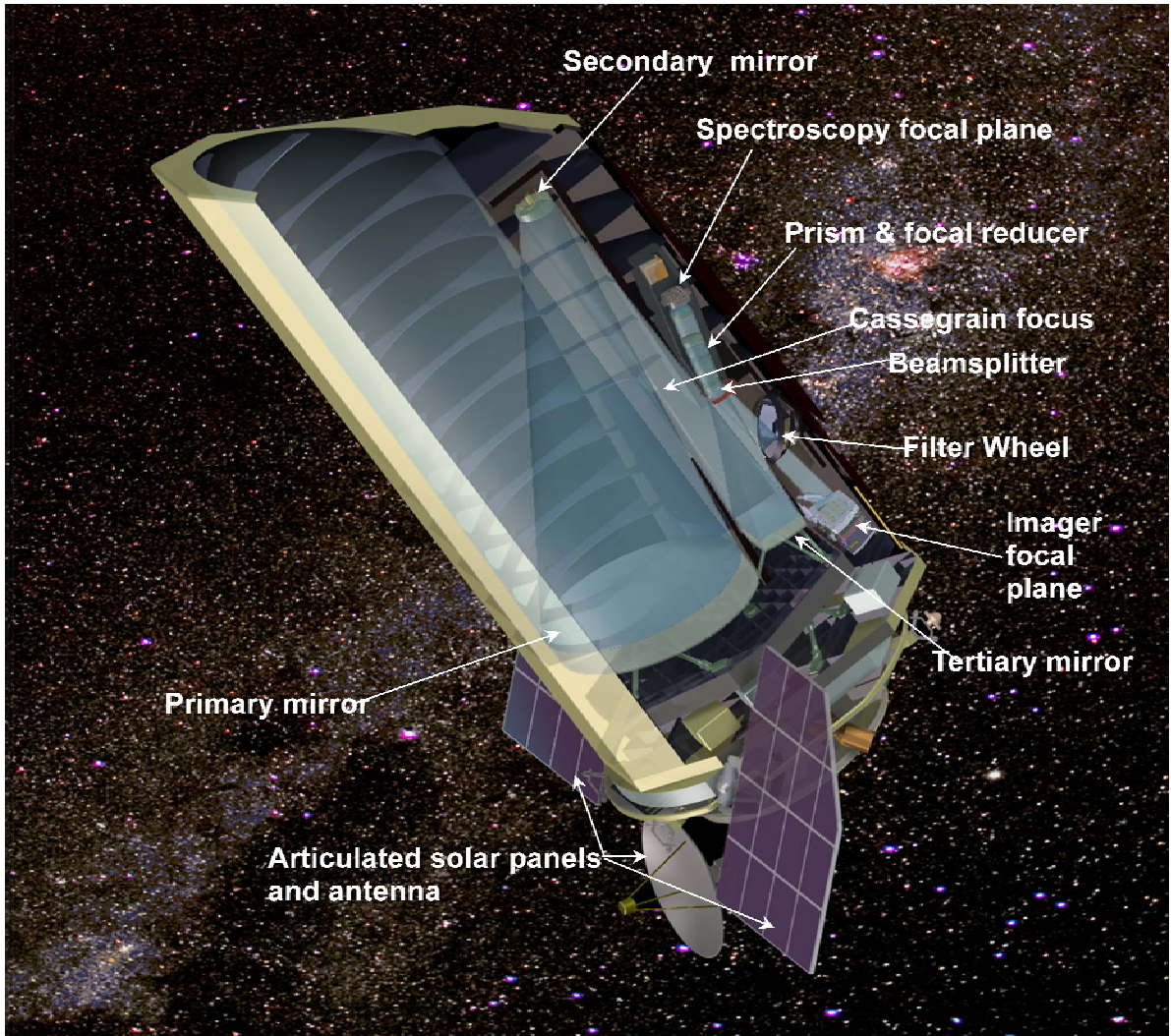

Figure 1: On-orbit configuration of WFIRST-A as described in the text. This system is an unobscured three-mirror anastigmat (TMA) with a dichroic beamsplitter at its exit pupil. The imager operates at the direct TMA f/15 speed, and accommodates a filter wheel as shown. The wide-field slitless spectrometer uses a prism in converging light and demagnifying optics to deliver spectra at f/6. At the Cassegrain focus, a pickoff and magnifying relay (not shown) feed a small-field IFU spectrometer. All focal planes are located on the cold side of the payload with passive cooling. The solar panels and K-band downlink antenna are fully articulated allowing extended observations of the Galactic Bulge for microlensing. Artwork by R.E.Lafever, LBNL.

An artist's concept of WFIRST-A is shown in Figure 1. The unobstructed three-mirror anastigmat telescope is arranged with its secondary and tertiary optics located on the shaded (cold) side of the payload where they are baffled against stray light and heat. The sensor packages are located there as well, for best passive cooling and convenient access for ground integration. The K-band antenna is fully articulated, and the solar panels are articulated about one axis. The outer baffle front cutoff angle is 45°, allowing sun-target angles > 45°.

In Figure 2 we illustrate the focal plane layouts on the sky, from Sholl et al. (2010, 2011), and list their pixel scales in arcseconds. A filter wheel can be installed into the imaging channel of any of these imager configurations without significant loss of performance (Lampton 2011).

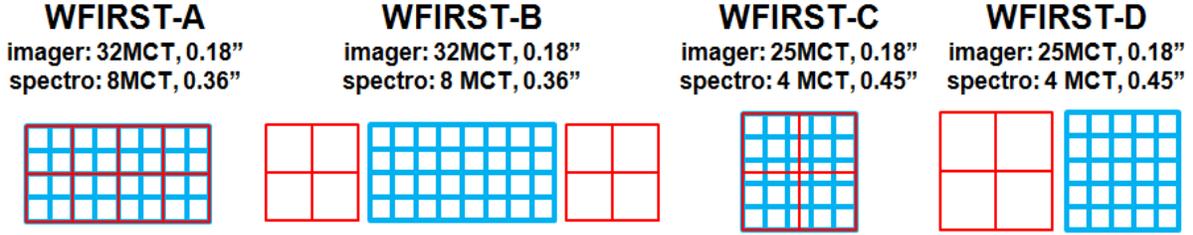

Figure 2: Sky fields for the four WFIRST configurations explored here, from Sholl et al. (2011). Red: spectrometer field; blue: imager field. Sky fields overlap in dichroic beamsplitter configurations A and C. Pixel scales shown are for 2K x 2K HgCdTe sensors with 18μm pixels. The compact Cassegrain layouts of A and C permit them to use a 1.5m primary mirror and fit within an EELV 4m launch fairing.

Survey rate is the single most important parameter for each science objective. The background limited survey rate reaching a given source flux with a required continuum signal to noise ratio SNR with a bandwidth Δλ and background (chiefly zodiacal) continuum flux B, is:

$$\text{Survey Rate} = \frac{\text{Flux}^2 \cdot \Delta\lambda}{4 \cdot \text{SNR}^2 \cdot B} \cdot \frac{\Omega_{fov} \cdot \text{AE}}{\Omega_{halfenergy}}$$

Here, Flux represents the photon continuum flux of the faintest required target, $\Omega_{fov}$ is the field of view, AE is the product of light gathering area and all efficiencies, and $\Omega_{halfenergy}$ is the solid angle of the image blur patch that contains half the target energy. A related formula applies for line emitter surveys:

$$\text{Survey Rate} = \frac{\text{Flux}^2}{4 \cdot \text{SNR}^2 \cdot B \cdot \Delta\lambda} \cdot \frac{\Omega_{fov} \cdot \text{AE}}{\Omega_{halfenergy}}$$

Again Δλ refers to the entire bandpass, but now Flux refers to the target's line photon flux at threshold. These expressions emphasize the need for a wide field of view, a large light gathering power, and a small half-energy spot size.

In this report we estimate survey rates using the public JSIM mission calculator (Levi, 2009) which takes into account a number of additional practical considerations: (a) the effective point spread function is not a number but is a function of wavelength, galaxy size, optical performance, pixel size, and attitude stability; (b) galaxy size is not a number but is a distribution that depends on redshift; (c) the instrumental throughputs and point-spread functions depend on wavelength and have to be integrated over each waveband; (d) for shorter exposures and for supernova spectroscopy, detector noise contributions become significant and must be included; (e) mission time is extended by sensor readouts, attitude maneuvers, and calibration sequences; (f) the Zodiacal foreground is a function of ecliptic latitude, longitude and wavelength; (g) the sky background becomes increasingly cluttered by Galactic extinction and Galactic star populations as mid- and low-latitude fields are examined; and (h) even at high Galactic latitudes, a fraction of potential slitless spectroscopy targets will be lost as a consequence of interference from other objects --- a situation that is ameliorated by performing dithers as a series of roll and translation maneuvers rather than translations alone.

In JSIM, the encircled energy (EE) function is evaluated from the Bessel transform of the system modulation transfer function (MTF), which itself has several contributors. Following Schroeder (2000) we normalize angles and angular frequencies to the angular cutoff frequency of the pupil given by $f_C = D/\lambda$ with $z = f/f_C$ and $w = \theta f_C$, and compose the MTF from five factors:

$$EE(w) = 2\pi w \cdot \int_0^1 \text{MTF}(z) \cdot J_1(2\pi w z) \cdot dz$$

$$MTF(z) = Tgal(z) \cdot Tacs(z) \cdot Tdiffract(z) \cdot Taberr(z) \cdot Tpixel(z)$$

The target galaxies are assumed to follow exponential profiles, for which *Tgal(z)* has a functional form $(1+kz^2)^{-1.5}$ where $k = (3.744 \cdot \text{Ree} \cdot D/\lambda)^2$ where Ree=half encircled energy radius and again z = normalized angular frequency. The attitude control system jitter is assumed to contribute a Gaussian broadening and hence has a Gaussian *Tacs(z)* factor as per Schroeder's eq 11.1.18. For the optical diffraction by a centrally obscured circular pupil, we adopt the usual formulism described by Wetherell (1980) for *Tdiffract(z)*. For the optical aberrations, we started with the field-averaged theoretically optimized wavefront error (WFE) from each WFIRST optical train from Sholl (2010, 2011) or from Sanz (2010), and doubled these to account for additional WFE arising from manufacturing, metrology, and launch induced misalignments. We then distributed that WFE into the spatial frequencies according to the Fischer et al (2000) exponential Hopkins ratio model with a wavefront correlation of 0.333 (see also Olson 2008) to obtain an aberration MTF contributor *Taberr(z)*. The *Tpixel(z)* factor was evaluated for square pixels using the azimuthally averaged broadening expression of Schroeder, eq.17.1.5.

The weak gravitational lensing survey requires that WFIRST determine shears for $\sim 10^9$ galaxies, provide accurate photometry to contribute usefully towards groundbased photo-z estimates, and obtain a spectroscopic sample to calibrate the photo-z estimates. To accomplish this for a 20,000 square degree survey field will require reaching a surface density $\sim 30$ galaxies per square arcminute and a threshold ABmag $\sim 25$ (Leauthaud et al., 2008; Jouvel et al., 2009) with an individual galaxy ellipticity error <15%, and a threshold signal to noise ratio, SNR, of 20. This SNR is driven both by ellipticity error and photometric error requirements.

The basis of our survey rate calculation is to model the imager wavelength bandpass as a single 30% wide band centered on 800 nm. Focal planes having 25 to 32 MCT arrays deliver instantaneous survey fields of 0.26 to 0.32 sq deg, as listed in Table 2. For pixel subsampling (Bernstein 2002; Bernstein and Jarvis 2002), our estimated survey rates assume four dithers per field separated by 40 sec dead time for sensor readout and attitude maneuvering. To deal with chromatic shear and to provide photo-z information it will be useful to subdivide each exposure into adjoining wavebands, an option not explored here. Depending on aperture, each field will require a total exposure of ~2 ksec. Approximately $4 \times 10^4$ image fields will constitute the 10,000 square degree weak lensing survey.

Our results are listed in Table 3. Compared to the JDEM-Ω payload, the WFIRST configurations show significantly improved weak lensing survey rate. Although their apertures, pixel scales, and focal plane fields of view are closely comparable, the significant difference is WFIRST's unobstructed aperture which leads to a more compact point spread function and therefore shorter exposure times reaching a given signal to noise ratio, and a higher density of resolved galaxies. These two benefits are particularly clear in the direct comparison of Euclid-Plus to Euclid. We find that both Euclid-Plus and the WFIRST options deliver twice the survey speed and improved galaxy count compared to JDEM-Ω and Euclid. In particular note the increased density of resolved galaxies. Remarkably, WFIRST-B and D accomplish this with telescopes smaller than the 1.5m aperture of JDEM-Ω.

A useful sample of spectroscopic redshifts for calibrating the weak lensing photo-z redshifts could be obtained using WFIRST's supernova spectrometer. This instrument could be either an integral-field IFU (10" x 10") or slit (100"x1") spectrometer. Either field is large enough that, during routine survey operations, statistically ~1 galaxy will fall into each spectrometer field, thereby delivering $\sim 4\times10^4$ random (unselected) galaxy spectra with no increase in time allocation since it would operate parasitically during the wide field surveys. Alternatively a smaller field SNS could be used in a targeted mode. These options have not been evaluated here.

Table 3
Weak Lensing Survey Results

|  | WF-A | WF-B | WF-C | WF-D | JDEM-Ω | Euclid | Euclid+ |
|---|---|---|---|---|---|---|---|
| SurveyRate, 25mag:  sqdeg/yr | 5578 | 4067 | 4616 | 3187 | 2190 | 2640 | 6592 |
| Galaxies per sq arcminute | 42 | 42 | 42 | 42 | 32 | 29 | 42 |

Note—For all missions we set exposure times to deliver a signal to noise ratio of 20 on ABmag=25 targets, for one waveband 30% wide at 0.8μm. Zodiacal flux is assumed to be the average of the darkest 10,000 square degree sky. Four dithers per field are assumed.

To conduct a baryon acoustic oscillation survey, WFIRST is expected to obtain spectra of $\sim 2\times10^8$ galaxies and determine reliable redshifts for $\sim 10^8$ of them over the redshift range $1.3<z<2$ based on the Hα emission line. The Decadal Survey recommended $1<z<2$ but expected advances in ground based spectroscopy programs will close this window appreciably; furthermore the JDEM ISWG (2010) recommended $1.3<z<2.0$. We have based our survey rate predictions (Table 4) on the galaxy luminosity function described by Ilbert et al. (2005) and the redshift distribution of Ilbert et al. (2006), and the need to achieve a minimum detectable flux of $2\times10^{-16}$ erg/cm$^2$sec for 66% of the emission line galaxies on the Hα line. Four roll dithers will be sufficient to disambiguate the prism spectra at this level (Kent, S., private communication 2011). For Euclid we have adopted a grism efficiency of 65% relative to the prisms in the other missions listed. We note that for slitless spectroscopy, the noise level on faint targets is set by the Zodiacal continuum intensity, which is relatively low in the 1.5-2.0μm band compared to shorter wavelengths. For this reason the restricted spectroscopy waveband (and restricted Hα redshift range) of our A and C options provides an improved survey rate compared to the others.

Table 4
Baryon Acoustic Oscillations Survey Rate Results

|  | WF-A | WF-B | WF-C | WF-D | JDEM-Ω | Euclid | Euclid+ |
|---|---|---|---|---|---|---|---|
| Survey Rate, sqdeg/yr, 0.7<z<2.0 | * | 4038 | * | 3026 | 4014 | 1863 | 4737 |
| Survey Rate, sqdeg/yr, 1.3<z<2.0 | 11223 | 7997 | 8469 | 6165 | 8331 | 3873 | 9251 |

Note—All missions including Euclid were evaluated for a minimum detectable Hα line flux of $2\times10^{-16}$ erg/cm$^2$sec with 66% success rate using 4 roll dithers.  *The WF-A and WF-C spectrometer variants span only 1.5<λ<2.0μm, hence are limited to Hα 1.3<z<2.0.

For exoplanet microlensing no simple survey rate expression describes the science yield owing to the complex background/foreground structure of the Galactic Bulge.  Rather than implement a lensing simulator, we have instead adopted a figure of merit based on comparing WFIRST's parameters to those of the well-studied Microlensing Planet Finder (MPF) mission (Bennett et al., 2009, and references therein).  The scaling parameters we adopted are field of view, area-efficiency product, bandwidth, center wavelength, observing efficiency, and the inverse of the half-energy diameter on the focal plane.  The WFIRST examples (Table 5) emerge closely comparable to MPF owing to our larger and unobscured telescope apertures, which helps tighten the point spread function, even though our WFIRST imagers have less field of view than MPF.  In comparison JDEM-Ω suffers from having a non-articulated solar panel that restricts its annual dwell on the Bulge to two periods centered on the equinoxes.  For Euclid we assume that exoplanet lensing measurements will be done with its Near Infrared Photometer H band imager.  Euclid's payload configuration (Wallner et al. 2010) has a nonarticulated solar panel which like JDEM-Ω limits its stay time on the Bulge. Its payload length is constrained by its Soyuz ST 2-1B launcher (Wallner et al. 2010), and Euclid therefore lacks a strongly angled forward extension of its outer baffle that would permit extended stays on the Bulge.

Table 5
Exoplanet Discovery Rate Results

|  | WF-A | WF-B | WF-C | WF-D | JDEM-Ω | Euclid | Euclid+ |
|---|---|---|---|---|---|---|---|
| Relative exoplanet discovery rate | 1.43 | 1.46 | 1.12 | 1.14 | 0.21 | 0.48 | 0.85 |

Note— Exoplanet discovery rates relative to Microlensing Planet Finder, scaled according to imager sky area, annual time available on the Bulge, and other factors (see text).

For supernova follow-up spectroscopy, we assume that a robust ground-based program and/or WFIRST imager data (JDEM ISWG recommendation) will serve to generate supernova triggers. WFIRST will deliver shallow spectra at a series points on each light curve, plus one deep spectrum taken near maximum light.  The shallow spectra permit synthetic photometry without errors from assigning K-corrections, while the deep spectrum is sufficient to clearly determine the line ratios that are most useful in typing SNe (e.g. SiII strong; HeI absent) for 0<z<1.4.  The spectrometer should deliver a resolving power > 70 and span 0.4<λ<2.0μm.  It can be accommodated in WFIRST by intercepting a small portion of the field at Cassegrain focus, which in all four configurations lies upstream of wide field imager and spectrometer.  There, a 10" field is smaller than 1mm and, at f/6, the relay optic and spectrometer will be compact.  We have assumed that each payload uses an integral field unit (IFU) ahead of its spectrometer in estimating the follow-up rates listed in Table 6.

Table 6
Supernova Survey Rate Results

|  | WF-A | WF-B | WF-C | WF-D | JDEM-Ω | Euclid | Euclid+ |
|---|---|---|---|---|---|---|---|
| SNe followed per year | 2130 | 1531 | 2129 | 1565 | See text | n/a | n/a |
| Figure of Merit, per year | 120 | 112 | 120 | 112 | | | |

Note— We assume that the four WFIRSTs use IFUs, and that exposure times are chosen to give a peak continuum signal/noise = 15 with a resolution of 100 per pixel, for SNe uniformly distributed out to z=1.4. In addition, sixteen shallow spectra are taken with signal/noise = 4 per pixel for a resolution of 30. Figure of merit is described in the text.

An implementation decision to be taken is the choice between a slit type spectrometer and a two-dimensional integral field unit (IFU) spectrometer. The IFU offers three clear benefits to the mission science: full capture of the SN host galaxy, freedom in choice of roll angle of observation, and improved attitude control tolerance. Although the slit spectrometer has greater flight heritage, we believe that the advantages of the IFU warrant its adoption for WFIRST, particularly in view of the extensive IFU flight qualification program pursued by the French Space Agency CNES (Ealet et al., 2006; Aumeunier et al., 2008; Cerna et al., 2008; Prieto et al., 2008; Cerna et al., 2010).

JDEM-Ω baselined a slitless spectrometer channel responsible for meeting the supernova spectroscopy requirement. In JDEM-Ω spectroscopy, the zodiacal background per pixel-resolution element is ~240 times higher than if an IFU were used, requiring ~240x longer exposure times when zodiacal-noise limited. Only a couple of SNe within a week of peak brightness are expected within the JDEM-Ω field for a given exposure. JDEM-Ω therefore does not compete well compared to missions with IFU-based or slit-based spectroscopy.

We note here that multi-epoch spectroscopy has a practical advantage over the Astier et al. (2010) plan with seven broad band filters: with the continuous spectrum available there is no need to apply band-dependent K-corrections. Multi-epoch spectroscopy identifies both the light curve shape and the evolution of spectral features, for SN classification and subtyping.

For the supernova survey figures of merit (FoM), we fit the cosmological parameters $\Omega_M$, $w_0$, $w_a$, and $\mathcal{M}$ assuming a flat universe, and calculate FoM defined as $1/\sqrt{(\det(C_{w0,wa}))}$, for an input flat $\Lambda$CDM cosmology with $\Omega_M$ =0.28. We supplement the WFIRST data with 300 externally-measured supernovae at z~0.05 and a Planck prior. Each supernova has a 0.15 mag random scatter from intrinsic dispersion and measurement uncertainty. For each survey, the WFIRST supernovae are uniformly distributed from 0.2<z<1.4. Our systematic error model, suggested by Linder & Huterer (2003), sets a floor for each 0.1 redshift bin with $\sigma_{sys}$=0.02(1+$z_{bin}$)/2.4. The resulting FoMs are given in Table 6 and demonstrate the supernovae and Planck alone provide more than a factor of two improvement compared to the combined DETF Stage II results from all dark energy probes (Albrecht et al., 2006).

## 4.  Specific Payload Features

The fact that a wide-field imager, a wide-field slitless prism spectrometer, and a narrow-field IFU spectrometer are simultaneously available in these WFIRST payload configurations will allow a highly productive joint survey to be conducted, yielding WL and BAO products and a representative sample of galaxy spectroscopic redshifts during the same mission phase. This simplification will materially benefit overall mission productivity. The individual science objectives can interleave into a mission in many ways ranging from highly parallel to fully serial.

Unlike missions like HST and JWST, a survey payload must deliver its good image quality over a large field of view. The usual limitations on mass, volume, cost, schedule argue strongly for adopting the smallest telescope that can deliver the required performance. Traditional space telescopes are centrally obstructed by their secondary mirrors and central stray light shields. These impose no serious performance impact provided that the blockage is small compared to the primary mirror diameter. However in a wide field instrument, the secondary mirror and its baffle must grow to accommodate the field, and the diffraction pattern suffers. As shown by Lampton et al. 2010, this diffraction pattern can seriously enlarge the half energy radius and hence degrade the survey rate, compared to an unobstructed pupil. Such an obstructed pupil loses 25% of the incident flux via direct blockage, and another 25% in peak flux owing to added diffraction loss to the surrounding rings. In Figure 3 below we show the growth in half light radius (EE50 radius) with increasing wavelength for four alternative telescopes.

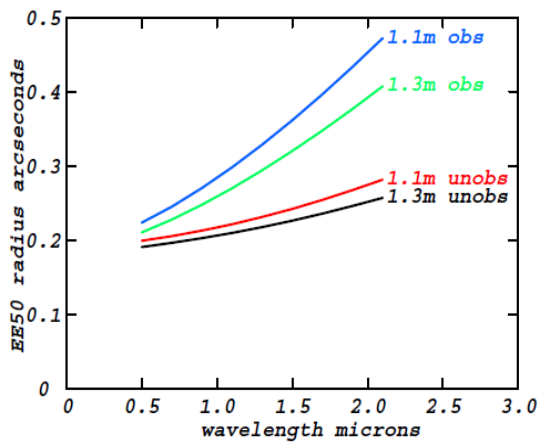

Figure 3: A system's half encircled light radius (EE50) is determined by a number of factors including aberrations, optical manufacturing errors, alignment, pixellization, and attitude control system jitter. Here we plot the EE50 radius for four examples: 1.1 and 1.3m apertures, unobscured and 50% linearly obscured. At increasingly long wavelengths, where signal-to-noise ratio is most demanding, diffraction dominates the other blur contributors, particularly for smaller telescope apertures. To reach the larger redshift targets it is important to minimize the diffraction, which unobstructed apertures accomplish. Background limited exposure times go as the square of the EE50 radius. After Lampton et al. (2010).

A second advantage of an unobscured pupil is the freedom from diffraction spikes originating in the pupil blockage caused by the secondary mirror support structure. These spikes complicate the determination of instrumental shear owing to their strongly chromatic nature and the concomitant errors in subtracting them from wideband images.

Although unobscured three-mirror anastigmat telescopes date back to Cook (1979) and Korsch (1980), most wide field telescope designs continue to use the centrally obscured Korsch (1977) annular and full-field TMAs, or the centrally obscured eccentric field Williams (1979) TMA. The traditional preference for axisymmetric primary mirrors is that the net aspheric departure, and hence the figuring time, is smaller. With contemporary computer-driven grinding and figuring, however, this economy has faded: indeed unobstructed apertures are now becoming popular for remote sensing applications owing to their improved image contrast per dollar spent.

Traditional space observatories (HST, JWST) employ a focal telescope to deliver a sky image to a focal plane, which is divided up into field regions dedicated to each onboard instrument. Each such instrument then can image directly or recollimate and refocus to conduct spectroscopy and/or imaging at its unique required focal length. In contrast, the dual-focal length concept of Sholl et al. (2010; 2011) yields a significant mission simplification for WFIRST where both imaging and spectroscopy are required.

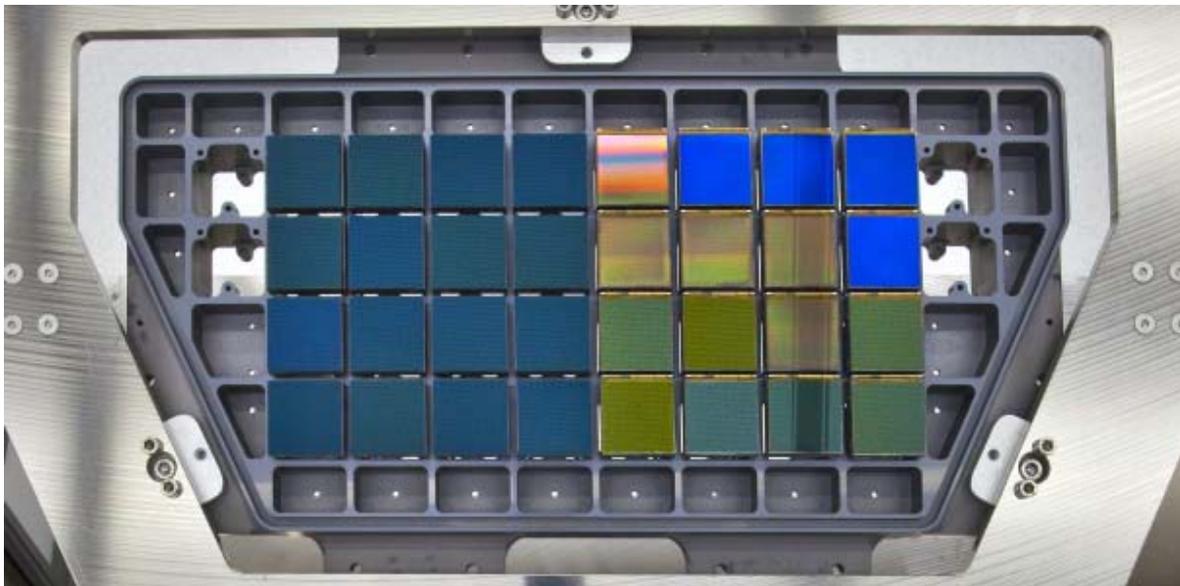

Figure 4: A 32-sensor imager focal plane accommodates modular Si CCDs and/or HgCdTe SCAs, with optional individual bandpass filters (Jelinsky et al. 2011; Besuner et al. 2011). The focal plane metering structure (gray) is silicon carbide. Here, the CCDs are 3512 x 3512 pixels on fully depleted p-channel Si (Bebek et., 2004; Baltay et al. 2010). HgCdTe SCAs are Teledyne H2RG® 2048x2048 pixels (Lorenzon et al. 2008; Schubnell et al. 2008, 2010). Each sensor includes low-power cryogenic electronics, yielding a photons-to-digital module. Four outrigger locations are available for reference photodiodes and/or star guiders. This assembly is shown in its qualification chamber mount during performance characterization, thermal cycling, and electromagnetic compatibility tests.

The combined weak lensing and exoplanet requirements impose a fine pixel scale (~0.18") and a large field of view (~0.3 sq deg) for the imager. These in turn drive the need for $> 10^8$ pixels in the imager, or >25 SCAs in a 2K x 2K format. DOE has sponsored the development and flight qualification of fully modular focal planes of the necessary size (see Figure 4) complete with integral low-power front end electronics (Lorenzon et al. 2008; Schubnell et al. 2008, 2010; Baltay et al. 2010; Besuner et al. 2010; Jelinsky et al. 2011; Besuner et al. 2011). The basic design uses SiC for the cryogenic metering platform and sensor mounts. To support its cryogenic qualification testing, an optical stability metering instrument has been developed (Edelstein et al. 2011). The focal plane provides additional integral mounting locations for optional guiders and for a complement of photometric diode sensors that assist flat field calibration (Scarpine et al., 2010; Baptista & Mufson 2010). The ratio of active sensor span to grid spacing is 4:5. By moving the sky field in diagonal steps whose x and y lengths are ¼ the sensor active area size, each group of five such steps yields a filled uniform depth survey. The large pixel count and the many dither steps per field deliver a useful tolerance to hot and dead pixels in the array. The sensor modules can be individually fitted with bandpass filters, offering an alternative to using a filter wheel: the color information could be generated as part of the shear mapping strategy.

For exoplanet microlensing survey work it is highly desirable to provide for frequent repeat visits (~ 1ksec cadence) to a number of fields near the Galactic Bulge, and to sustain this visit pattern for an extended period of time considerably longer than the 20-50 day durations of typical lensing events. These requirements have three mission impacts. First, the optical system will need to baffled in such a way as to permit observing targets as close as 50° to the sun; our recommended configurations share a long forward baffle extent and a 45° cutoff angle to permit this targetting. Figure 5 illustrates the visibility of the Bulge over the course of one year: the green zone shows that the sun-target angle exceeds 50° for nearly 9 months of the year. Second, the solar panels must be fully articulated to maintain a positive power margin throughout the needed range of sun-target angles. Third, to maximize time-on-target, it is desirable to use a fully articulated K-band antenna so that downlink can proceed during data taking.

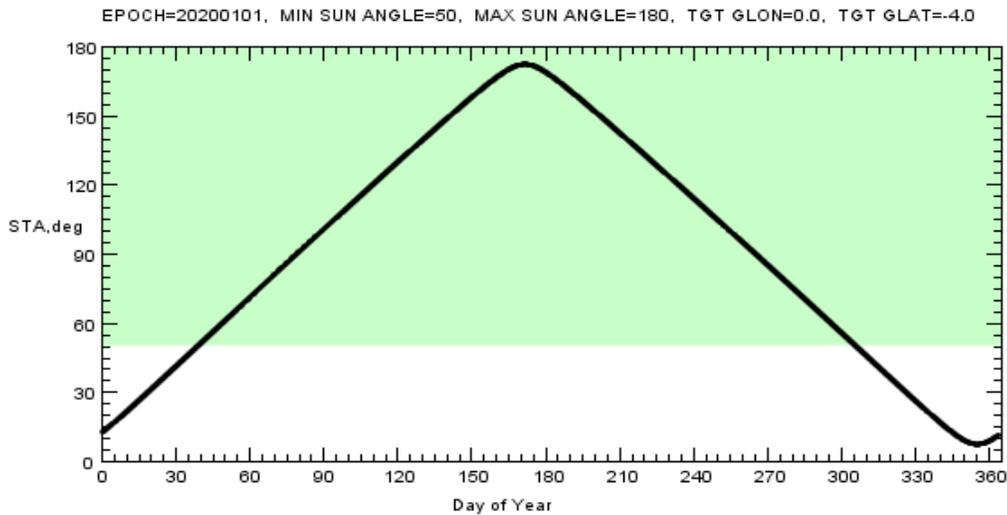

Figure 5: Sun-target angle STA for Baade's window as a function of day of year. The green band is the range of STAs that is accessible with the 45° front sunshade angle.

Driven by the requirement for maximizing dark sky time, dark energy mission planning (e.g. Euclid; JDEM) is usually based on observing from a halo orbit about the Sun-Earth L-2 Lagrange point. Such a location is free from eclipses and has essentially no interfering optical glare from Earth or Moon, and need not deal with the geomagnetically trapped outer-belt radiation environment. In contrast, explanet microlensing concepts (GEST: Bennett et al. 2003; MPF: Bennett et al. 2004, 2009) baseline the use of an inclined geosynchronous orbit which – although it poses a heavy radiation environment -- provides a continuous downlink capability to a single ground station, ideal for handling a sustained high data rate economically.

The NASA-GSFC Mission Development Laboratory studied downlink options for SNAP/JDEM during 2008. In particular they explored four Ka-band high-rate data return alternatives from a spacecraft located at L-2 using ground facilities that were anticipated to come online in this decade. We summarize those studies in Table 7.

Table 7
Alternative K-band Downlink Options From Earth-Sun L-2 Orbit Locations

|  | White Sands | DSN Gladstone | DSN Madrid | DSN Canberra |
|---|---|---|---|---|
| Data rate, Mb/s | 50 | 150 | 150 | 150 |
| S/C antenna, m | 1.2 | 1.2 | 1.2 | 1.2 |
| XMT power, W | 35 | 40 | 40 | 40 |
| Link margin, dB | 2.10 | 8.06 | 6.51 | 3.37 |

For comparison, the WFIRST payloads described here average ~ 50Mb/sec for 3h/day during combined WL+BAO surveys and ~60Mb/sec for 24h/day while surveying exoplanet fields. A new study should be conducted to reflect current and future ground capabilities and to identify a combination of data compression (e.g. Bernstein et al. 2010; Pence et al. 2010), onboard storage, region-of-interest processing on subsampled maps, and multiple-ground-station support that best meets the combined needs of WFIRST and the many other Deep Space Network users.

## 5. Conclusions

We have defined four potential WFIRST payloads that accomplish the Decadal Survey's objectives and that fully utilize recent optical and payload advances. These four alternatives deliver significantly higher performance than the Joint Dark Energy Mission JDEM-$\Omega$ concept considered by the Decadal Survey. We also compared these configurations to JDEM-$\Omega$ and to the European Space Agency's Euclid mission, and to an alternative Euclid configuration making use of the lessons learned here that delivers performance approaching that of WFIRST. We find that the unobstructed pupil is a major benefit to weak lensing owing to its tighter point spread function, improved signal to noise, and higher resolved galaxy count, allowing (for instance) WFIRST-A to deliver 2.5 times the survey rate of JDEM-$\Omega$, and an improved density of resolved galaxies, with a primary mirror of equal size. Similarly Euclid-Plus offers a factor of 2.5 higher weak lensing survey rate than Euclid. For BAO surveys, the zodiacal continuum in the $1.5 < \lambda < 2.0 \mu m$ band is substantially fainter than in the broader 1-2$\mu m$ band. This fact gives WFIRST-A an advantage in signal to noise ratio and hence in survey speed compared to wider-span spectrometers. For example, WFIRST-A surveys 2.8 times faster than JDEM-$\Omega$ and six times faster than Euclid; Euclid-Plus offers more than twice the survey speed of Euclid. To address the supernova requirement the WFIRST payloads described here can follow up >1500 SNe per year of dedicated time with multi epoch spectroscopy. For exoplanets, these WFIRST payloads can deliver detection rates comparable to the dedicated MPF mission without significantly compromising the other objectives.

## Acknowledgments

The authors gratefully acknowledge support by the Director, Office of Science, of the U.S. Department of Energy under Contract No. DE-AC02-05CH11231. We thank Henry Heetderks for his many useful contributions to this effort. The authors would also like to thank Alexie Leauthaud and Nick Mostek for providing the weak lensing and BAO parameterizations from the COSMOS dataset that are used in JSIM.

# References

Albrecht, A., et al., "Report of the Dark Energy Task Force," Report to AAAC and HEPAP advising DOE, NASA, and NSF; arXiv 0609591 (2006).

Astier, P., Guy, J., Pain, R., and Balland, C., "Dark energy constraints from a space-based supernova survey," arXiv 1010.0509 (2010).

Aumeunier M-H., Ealet, A., Prieto, E., and Cerna, C., "First results for the spectrophotometry calibration of the SNAP spectrograph demonstrator in the visible range," Proc SPIE 7010 #3N (2008).

Baltay, C., Emmet, W., Rabinowitz, D., Szymkowiak, A., Bebek, C., Eames, J., Karcher, A., Kolbe, W., Roe, N., Derwent, P., Diehl, H., Estrada, J., and Howell, J., "Space qualified abuttable packaging for LBNL p-channel CCDs, Part I," Proc. SPIE 7742-2E (2010).

Baptista, B. and Mufson, S., "Radiation hardness studies of InGaAs photodiodes at 30, 52, & 98 MeV and fluencies to $10^{10}$ protons/cm2," Proc. SPIE, 7742 #25 (2010).

Bebek, C., J., et al., "Fully depleted back-illuminated p-channel CCD development," Proc. SPIE 5167, #50 (2004).

Bennett, D. P., et al., "The Galactic Exoplanet Survey Telescope (GEST)," Proc SPIE v.4854 (2003).

Bennett, D. P., et al., "The Microlensing Planet Finder: completing the census of extrasolar planets in the Milky Way," Proc SPIE v.5487 (2004).

Bennett, D. P., et al., "A census of exoplanets in orbits beyond 0.5 AU via space-based microlensing," White Paper for the Astro2010 PSF Science Frontier Panel, arXiv 0902.3000 (2009).

Bernstein, G. "Advanced exposure-time calculations: undersampling, dithering, cosmic rays, astrometry, and ellipticities," PASP v.114, 98-111 (2002).

Bernstein, G., and Jarvis, M. "Shapes and shears, stars and smears: optimal measurements for weak lensing," Astron.J., v.123, 583-618 (2002).

Bernstein, G., Bebek, C., Rhodes, J., Stoughton, C., Vanderveld, R. A. & Yeh, P. "Noise and Bias In Square-Root Compression Schemes," PASP, 122, 336 (2010).

Besuner, R., Baltay, C., Diehl, H., Emmet, W., Harris, S., Jelinsky, P., Krider, J., Rabinowitz, D., and Roe, N., "Space qualified abuttable packaging for LBNL p-channel CCDs, Part II," Proc SPIE 7742-0H (2010).

Besuner, R.,  Bebek, C.J., Haller, G.M., Harris, S., Hart, P.A., Heetderks, H. D., Jelinsky, P.N., Lampton, M.L., Levi, M.E., Maldonado, S.E., Roe, N.A., Roodman, A., and Sapozhnikov, L., "A 260 megapixel visible/NIR mixed technology focal plane for space," Proc SPIE, 8155A-12, San Diego (2011).

Cerna, C., Aumeunier, M-H., Prieto, E., Ealet, A., Karst, P., Castera, A., Smadja, G., Soilly T., and Crouzet, P. E., "Setup and performances of the SNAP spectrograph demonstrator," Proc SPIE 7010 #1A (2008).

Cerna, C., Smadja, G., Ccastera, A., and Ealet, A., "Extraction of the frequency spectrum of the noise of a HAWAII 2RG NIR detector and impact on low-flux measurements," Proc SPIE 7742 #1J (2010).

Content, D., Dittman, M., Firth, B., Howard, J., Jackson, C., Lehan, J., Mentzell, J., Basquale, B., and Sholl, M., "Joint Dark Energy Mission optical design studies," Proc SPIE 7731-1D (2010).

Cook, L. G., "Three-mirror anastigmat used off-axis in aperture and field," Proc. SPIE v.183, 207-211 (1979).

Duvet, L., "Euclid Reference Payload Concept," SRE-PA /2010.030, European Space Agency, ftp://ftp.rssd.esa.int/pub/EUCLID-ITT/documents (2010).

Ealet, A., Prieto, E., Bonissent, A., Malina, R., Aumeunier, M-H., Cerna, C., Smadja, G., and Tilquin, A., "An integral field spectrograph for the SNAP Mission," Proc SPIE v.6265 #33 (2006).

Edelstein, J., Sirk, M., Hoff, M., and Jelinsky, P. "Cryogenic Focal Plane Flatness Measurement with Optical Zone Slope Tracking,"  Proc. SPIE 8155A-22, San Diego (2011).

Fischer, R. E., and Tadic-Galeb, B., *Optical System Design*, Ch.16, McGraw-Hill, NY (2000).

Gehrels, N., "The Joint Dark Energy Mission (JDEM)  Omega," arXiv:1008.4936  (2010).

Grange, R., Milliard, B., Kneib, J-P., and Ealet, A. "A simple optical design for a space dark energy mission,"  Proc. SPIE v.7731 #3H (2010).

Heetderks, H.D.,  "Launch fairing accommodation study for WFIRST payloads," JDEM Project Office Drawing No. 1029-A (7 Dec 2010).

Ilbert, O., et al., "The VIMOS-VLT deep survey: Evolution of the galaxy luminosity function up toe z=2 in first epoch data," Astron. & Astroph. v.439, 863-876, (2005).

Ilbert, O., et al., "Accurate photometric redshifts for the CFHT legacy survey calibrated using the VIMOS VLT deep survey," Astron. & Astroph. v.457, 841-856, (2006).

JDEM Interim Science Working Group (ISWG) Final Report to NASA and DOE (May 2010) http://wfirst.gsfc.nasa.gov/science/ISWG_Report.pdf

Jelinsky, P. N., Bebek, C. J., Besuner, R. W., Haller, G. M., Harris, S. E., Hart, P. A., Heetderks, H. D., Levi, M. E., Maldonado, S. E., Roe, N. A., Sapozhnikov, L., "The design and integration of multiple technology focal planes," American Astronomical Society 217[th] Meeting, Seattle, 433.02 (2011).

Jouvel, S., Kneib, J-P., Ilbert, O., Bernstein, G., Arnouts, S., Dahlen, T., Ealet, A., Milliard, B., Aussel, H., Capak, P., Koekemoer, A., Le Brun, V., McCracken, H., Salvato, M., and Scoville, N., "Designing future dark energy space missions: I. Building realistic galaxy spectro-photometric catalogs and their first applications," Astron. & Astroph. v.504, 359 (2009).

Korsch, D., "Anastigmatic three-mirror telescope," Applied Optics v.16#8, 2074-2077 (1977).

Korsch, D., "Design and optimization technique for three-mirror telescopes," Applied Optics v.19#21, pp.3640-3645 (1980).

Lampton, M., Sholl, M., and Levi, M., "Off-axis telescopes for dark energy investigations," Proc. SPIE 7731-1G (2010).

Lampton, M., "Impact of installing a filter wheel into the WFIRST Imagers," JDEM Project Office Drawing No. 1165-A (2011).

Levi, M. "JDEM Mission Calculator" (JSIM) http://jdem.lbl.gov/public/etc/jsim.html (2009).

Levi, M.E., Lampton, M., and Sholl, M. "Wide Field InfraRed Survey Telescope Science Yield," 217[th] AAS Meeting, Seattle, 433.03 (2011).

Linder, E.V., and Huterer, D.,"Importance of supernovae at z>1.5 to probe dark energy," Phys. Rev. D 67, 081303 (2003).

Lorenzon, W., Newman, R., Schubnell, M., Tarle, G., and Weaverdyck, C., "Count rate dependent non-linearity and pixel size variations in 1.7 micron cut-off detectors," Proc. SPIE 7021 #10V (2008).

National Research Council (NRC), Committee for a Decadal Survey of Astronomy and Astrophysics, "New Worlds, New Horizons in Astronomy and Astrophysics," National Academies Press (2010).

Olson, C., "Lens performance budgeting using the Hopkins ratio," Optics Photonics News v.12 p.12-15 (2008).

Pence, W. D., White, R. L., and Seaman, R., "Optimal compression of floating-point astronomical images without significant loss of information," arXiv 1007.1179 (2010).

Prieto, E., Ealet, A., Milliard, B., Aumeunier, M-H., Bonissent, A., Cerna, C., Crouzet, P-E., Karst, P., Kneib, J-P., Malina, R., Pamplona, T., Rossin, C., Smadja, G., and Vives, S., "An integral field spectrograph for SNAP," Proc. SPIE 7010 #19 (2008).

Sanz, I. E. "Euclid RD4 Reference Optical Design 1.2," Fig. 18, SRE-PA/2010.031, European Space Agency (2010).

Scarpine, V., Kent, S., Deustua, S., Sholl, M., Mufson, S., Ott, M., Wiesner, M., and Baptista, B., "The ring of fire: an internal illumination system for detector sensitivity and filter bandpass characterization," Proc. SPIE 7731-3E (2010).

Schroeder, D. J., *Astronomical Optics 2nd Edition*, section 11.1, Academic Press, San Diego, CA, (2000).

Schubnell, M., Brown, M. G., Karabina, A., Lorenzon, W., Mostek, N., Mufson, S., Tarle, G., and Weaverdyck, C., "Precision Quantum Efficiency Mesurements onf 1.7 Micron Near-Infrared Devices," Proc. SPIE 7021 #10L (2008).

Schubnell, M., Biesiadzinski, T., Lorenzon, W., Newman, R., and Tarle, G., "Investigating reciprocity failure in 1.7 micron cutoff HgCdTe detectors," Proc. SPIE 7742 #1N (2010).

Sholl, M., Kaplan, M., and Lampton, M., "Three mirror anastigmat survey telescope optimization," Proc. SPIE 7010 (2008).

Sholl, M., Bernstein, G., Content, D., Dittman, M., Howard, J., Lampton, M., Lehan, J., Mentzell, J., and Woodruff, R., "Observatory conceptual development for the Joint Dark Energy Mission," Proc. SPIE 7436-2 (2009)

Sholl, M., Content, D., Lampton, M., Lehan, J., and Levi, M., "Widefield spectroscopy and imaging at two plate scales with a focal three mirror anastigmat," Proc. SPIE 7731-1F (2010).

Sholl, M., Lampton, M., and Levi, M.E., "A Practical Implementation of the Wide Field InfraRed Survey Telescope WFIRST," 217th AAS, Seattle, 433.04 (2011).

Wallner, O., Ergenzinger, K., Tuttle, S., Vaillon, L., and Johann, U. "Euclid Mission Design," Proc. Intl. Conf. Space Optics ICSO 2010, Rhodes Greece (2010).

Wetherell, *Applied Optics and Optical Engineering*, v.8 Ch.6, Reading MA, Academic Press (1980).

Williams, S. G. L.,"On-axis three-mirror anastigmat with an offset field of view," Proc. SPIE v.183, pp.212-217 (1979).